\documentclass[12pt]{article}
\usepackage{amsmath,amsthm,amscd,amsfonts,amssymb}
\usepackage{latexsym}
\usepackage[usenames]{color}

\newcommand{\RR}{\mathbb R}

\newcommand{\ZZ}{\mathbb Z}
\newcommand{\EE}{\mathbb E}

\newcommand{\cc}{\mathcal C}
\newcommand{\dd}{\mathcal D}
\newcommand{\hh}{\mathcal H}

\newcommand{\NN}{\mathcal N}
\newcommand{\nn}{\mathcal N}

\newcommand{\WW}{\mathcal S}
\newtheorem{thm}{Theorem}[section]

\newtheorem{cor}[thm]{Corollary}

\newtheorem{rem}[thm]{Remark}
\newtheorem{hyp}[thm]{Hypothesis}
\newtheorem{exam}[thm]{Examples}

\newcommand\n{{\bf n}}
\renewcommand\k{{\bf k}}
\newcommand\m{{\bf m}}
\def\half{\frac{1}{2}}

\begin{document}
\author{M Krishna \\ Institute of Mathematical Sciences \\ Taramani Chennai 600 113 \\ India}
\title{Absolutely continuous spectrum and spectral transition for some continuous random operators}
\date{}
\maketitle
\begin{center}
{\it Dedicated to Barry Simon for his ${65}^{th}$ birthday.}
\end{center}
\begin{abstract} 
In this paper we consider two classes of random Hamiltonians on $L^2(\RR^d)$
one that imitates the lattice case and the other a Schr\"odinger
operator with non-decaying, non-sparse potential both of which exhibit
a.c. spectrum.  In the former case we also know the existence of dense
pure point spectrum for some disorder thus exhibiting spectral
transition valid for the Bethe lattice and expected for the Anderson model
in higher dimension.  

\end{abstract}

\section{Introduction and Main theorems}

In this paper we consider a two classes of random potentials and show
the absence of point spectrum for the corresponding random Schr\"odinger
operators for large energies.  We are motivated by the models
considered by Rodnianski-Schlag \cite{RodSch} and those by
Hislop-Kirsch-Krishna \cite{hkk}. 

Surprisingly the methods of proof in both the models are well known, one
being the use of commutators and the other wave operators.

Commutators have played a significant role in spectral and scattering theory with
the Kato-Putnam theorem (see Reed-Simon \cite{RS}) and the Mourre theory (see 
Mourre \cite{Mou} and Perry-Sigal-Simon)  \cite{PSS} )
addressing the presence of absolutely
continuous spectrum.  Positive commutators also have been used
in the spectral theory of random operators by Howland \cite{How} and by 
Combes-Hislop-Mourre \cite{CHM},  Krishna-Stollmann \cite{KriStol} 
even to show the continuity of density of states. 

On the other extreme non-zero commutators imply the
absence of point spectrum, an indirect fact  well known as the 'virial theorem'.
In the literature mostly this fact was used to conclude  
the absence of positive eigen values in the 
scattering theoretic models (see Kalf \cite{Kal}, Weidmann \cite{Wie}, Reed-Simon \cite{RS}
and Amrein-Anne Boutet de Monvel-Georgescu \cite{AmBerGeo} for example). 

A very general
discussion on the 'virial theorem' is given in Georgescu-Gerard \cite{GeoGer}
who give a collection of conditions under which the above theorem
is valid when $A$ is an unbounded self-adjoint operator. 
They also give an example where the theorem fails when $f$ is not
in the domain of $A$. 
 
This theorem is often used to show that there are no eigen values
in some set or there are no eigenvalues at all, see for example, 
Weidmann \cite{Wie}, Theorem VIII.59, Reed-Simon \cite{RS}, 
Proposition II.4, Mourre \cite{Mou}, Amrein-Anne Boutet de Monvel-Georgescu \cite{AmBerGeo}.  

We apply the 'virial' theorem to  models of random potentials 'living on large islands'
an extension of a class of models considered by  Rodnianski-Schlag \cite{RodSch}.  
As far as know this result is not known in the literature and includes 
random potentials which are neither 'decaying' nor are 'sparse' as we later
exhibit in the example \ref{exam1}.

Let $\beta \geq 0$ and let $r_\beta(x)$ be a positive function on
$\RR^d$ satisfying
$$
c_1 |x|^\beta \leq r_\beta(x) \leq c_2 |x|^\beta, ~~ \mathrm{for ~ some} ~ 0 < c_1 \leq c_2 <\infty.
$$
Let $\NN_{\beta, \gamma} $ be a discrete  subset of 
$\RR^d$ 
such that for points $x, y \in \NN_{\beta, \gamma}, ~ x \neq y$,  we have 
$$
\{w : |x - w | \leq \gamma r_\beta(x) \}  
\cap \{w : |y - w | \leq \gamma r_\beta(y) \} = \emptyset,  
$$
for some $0 < \gamma \leq 1.$

(It is easy to think of the case $r_\beta(x) = |x|^\beta$.)

Let $\{\omega_n, n \in \nn_{\beta,\gamma}\}$ be independent real valued random variables and let 
$\alpha \geq 0$. We define  random functions 
$V^\omega_{\beta, \gamma, \alpha}$ on $\RR^d$ as follows.
\begin{equation}\label{pot1}
V^\omega_{\beta,\gamma,\alpha}(x) = 
\sum_{n \in \NN_{1, \gamma}} \omega_n |n|^{-\alpha} \phi(\frac{x-n}{r_\beta(n)}).
\end{equation}
where $\phi$ is a smooth bump function supported in the unit ball in $\RR^d$ (
so that the $n$th  summand is a function centered at $n$ and supported
in a ball of radius $r_\beta(n)$ which is roughly $|n|^\beta$).
We will denote the operator on $L^2(\RR^d)$ of multiplication by the function 
$V^\omega_{\beta, \gamma, \alpha}$ by the same symbol.

\begin{thm}\label{thm3}
Consider the sets $\nn_{\beta, \gamma}$, i.i.d random variables $\{\omega_n,  n \in \nn_{\beta,\gamma}\}$
with compactly supported distribution $\mu$ and consider the  random Schr\"odinger operators
$$
H^\omega_{\beta,\gamma,\alpha} = -\Delta + V^\omega_{\beta,\gamma,\alpha}
$$
on $L^2(\RR^d)$.  
\begin{enumerate}
\item Let $\alpha, \beta \geq 0, ~ \alpha + \beta \geq 1$, 
then there is a $E_0 <\infty$ such that 
$$
\sigma_{pp}(H^\omega_{\beta,\gamma,\alpha}) \cap (E_0, \infty) = \emptyset.
$$
\item Suppose $\alpha + 2\beta \geq 2$,  then 
$$
\sigma_s(H^\omega_{\beta, \gamma, \alpha}) \cap (E_0, \infty) = \emptyset, ~ \mathrm{for ~ all} ~ \omega.
$$
\end{enumerate}
\end{thm}

\begin{rem}
\begin{enumerate}
\item The case $\alpha =0$ is very interesting especially for $d\geq 2$ 
as the randomness is neither decaying nor 
sparse!  The theorem  says that if we consider shallow independent potentials
on huge islands then there is no localization for large energies. This is a dimension
independent statement. 
\item In the case when $d \geq 2$ and $\beta > \half$, $\alpha = 3/4$, Rodnianski-Schlag \cite{RodSch}
showed the existence of modified wave operators for the pair $H^\omega_{\beta,\gamma,\alpha}, -\Delta$ and thus
showed that $\sigma_{ac}(H^\omega_{\beta,\gamma,\alpha}) = [0, \infty)$.  We consider weaker conditions on $V^\omega$
but also weaker conclusions.
\item In the above theorem all we need is that $[V^\omega_{\beta,\gamma,\alpha}, A]$ extends
to a bounded operator from $\WW(\RR^d)$, say, to $L^2(\RR^d)$, where
$A$ is the generator of dilation group given below.  
\item In the case $\beta =1$, the 'thickest' possible sets 
$\nn_{\beta, \gamma}$ are in some sense opposite of the Bethe
lattice.  The number of points $N(R)$ at a distance $R$ from the 
origin here grows logarithmically in $R$ asymptotically,  while on 
the Bethe lattice $N(R)$ grows exponentially.  
When $\beta$ varies from $1$ to $0$, the growth behaviour
of $N(R)$ changes from logarithmic to polynomial.
\end{enumerate}
\end{rem}

Taking the case $\alpha =0, \beta=1 $ in the above theorem we see that

\begin{cor}\label{cor2}
Let $0 < \gamma < 1$ and let $\phi$ be a smooth function supported in
a ball of radius $\gamma$ centred at the origin in $\RR^d$. Let
$\{\omega_n, n \in \nn_{1, \gamma}\}$ be i.i.d. random variables
distributed according to a compactly supported distribution $\mu$. 
Consider the random operators 
$$
H^\omega = -\Delta + V^\omega, ~ ~   
V^\omega(x) = \sum_{n \in \nn_{1,\gamma}} \omega_n \phi(\frac{x-n}{r_1(n)}), ~ \gamma < 1
$$
on $L^2(\RR^d)$.
Then there is a $E_0 < \infty$ such that the spectrum of $H^\omega$
in $(E_0, \infty)$ is purely absolutely continuous.
\end{cor}

\begin{rem}
Since $r_1(n) \approx |n|$, the
corollary gives non-decaying, non-sparse potentials with a.c. spectrum
and is also valid in one dimension. The potential configurations
consist of independent barriers or wells whose supports 
together 'cover' a fraction of $\RR^d$.  
\end{rem}

We want to make sure that there is spectrum in the region of energies
we are interested in and this is guaranteed by the following theorem.

\begin{thm}\label{cor1}
Consider the operators $H^\omega_{\beta, \gamma, \alpha}$ given in theorem \ref{thm3} with
$\alpha >0$.  Then
$$
\sigma_{ess}(H^\omega_{\beta, \gamma, \alpha}) = [0, \infty).
$$
Therefore the spectrum in $(-\infty, 0)$, if any, is discrete for each $\omega$.
If $\alpha =0$, then for every $E >0$, 
$$
\sigma(H^\omega_{\beta, \gamma, \alpha} ) \cap (E, \infty) \neq \emptyset.
$$
\end{thm}

The second model we consider comes from the paper of Hislop-Kirsch-Krishna 
\cite{hkk}.  Let $I$ be a set defined as in the appendix after equation \ref{eqn100} and $\Psi$ a multidimensional wavelet indexed by the set as in
hypothesis \ref{hyp1.1} and $\{\omega_n\}$ random variables satisfying hypothesis
\ref{hyp1.2}.  Let 
\begin{eqnarray*}
\Lambda & = & \{n \in \ZZ^d : |n_i| < \infty, ~ \mathrm{for ~ some} ~  i =1, \dots, d\} \\
I_\Lambda  &= & F\times \ZZ \times \Lambda \subset I  \\
P_{\n}  &= & |\Phi_\n\rangle\langle \Phi_\n|, ~ n \in I_\Lambda 
\end{eqnarray*}
and consider the operators
\begin{equation}\label{theothermodel}
H_\Lambda^\omega = -\Delta + \sum_{\n \in I_\Lambda} \omega_n P_\n.
\end{equation}

Combining with a theorem (theorem 3.5) of Hislop-Kirsch-Krishna \cite{hkk}  
we have

\begin{thm}\label{thm5}
Consider the operators $H_\Lambda^\omega$ given in equation 
(\ref{theothermodel}) such that the hypothesis \ref{hyp1.1} and 
\ref{hyp1.2} are satisfied.
Then 
\begin{enumerate}
\item $\sigma_{ac}(H_\Lambda^\omega) \supset [0, \infty), ~ \mathrm{for ~ all} ~\omega$.
\item There is a $E(\mu) < 0$ such that 
the essential spectrum of $H_\Lambda^\omega$ in $(-\infty, E(\mu))$ is 
non-empty and is pure point.
\end{enumerate} 
\end{thm}

\begin{rem}
We would like to point out a subtlety involved in the proof of (2) of the
above theorem.  The weak / intermediate disorder case of fractional
moment method of Aizenman \cite{Aiz} was used in the proof by
Hislop-Kirsch-Krishna \cite{hkk} in proving (2).  This proof considers
spectrum of the random operator in the resolvent set of the free
part and so gives
purity of the point spectrum even if one takes 
$V^\omega = \sum_{n \subset K} \omega_n P_n$, with
$P_n$'s mutually orthogonal rank one projections and
$\sum_{n \subset K} P_n \neq I$.  This proof should be
contrasted with the  method of Aizenman-Molchanov \cite{AM} which
(implicitly) requires $I - \sum P_n$ to be finite rank.
\end{rem}

\noindent {\bf Acknowledgement: } We thank Frederic Klopp for useful
correspondence.

\section{Proofs of the theorems}

We start with recollecting a more 'practical' version of the 'virial theorem' 
which is given  
in Proposition 7.2.10 of Amrein-Anne Boutet de Monvel-Georgescu \cite{AmBerGeo},
incorporating also the conditions from their theorem 6.2.10. 

\begin{thm}[Virial Theorem]\label{thm4}
Suppose $H, A$ is a pair of self adjoint operators on a separable Hilbert
space $\hh$ such that
\begin{enumerate}
\item there is a constant $c < \infty$ such that for all $f \in D(H) \cap D(A)$,$$
|\langle Hf, Af\rangle - \langle Af, Hf\rangle| \leq c (\|Hf\|^2 + \|f\|^2)
$$ 
and
\item for some $z \in \rho(H)$, the set
$$
\{f \in D(A) : R(z)f, R(\overline{z})f \in D(A) \}
$$
is a core for $D(A)$.
\end{enumerate}
Then,
$$
\langle Hf, Ag\rangle - \langle Af, Hg\rangle = 0.
$$
whenever $f,g$ are  eigen vectors of $H$ with the same eigen value.  
\end{thm}

{\noindent \bf Proof of Theorem \ref{thm3} :}

In the following we drop the indices $\alpha, \beta, \gamma$ on both $V^\omega_{\beta,\gamma,\alpha}$ and $H^\omega_{\beta,\gamma,\alpha}$
for ease of reading.  

To prove (i) we first note that since $V^\omega_{\beta,\gamma,\alpha}$ is a bounded operator, $D(H^\omega_{\beta,\gamma,\alpha})
= D(-\Delta)$.  
We consider the generator of dilation group $A = -i \sum_{i=1}^d \left( x_i\frac{\partial}{\partial x_i} + \half\right)$. It is well known that the Schwartz
space of rapidly decreasing functions $\WW(\RR^d)$ is a core for $A$
and the commutator of $A$ and $-\Delta$ is computed as  
$$
i[-\Delta, A] = -2\Delta
$$
on $\WW(\RR^d)$ and extends to $D(-\Delta)$.  
Let us set $\phi_n(x) = \phi( \frac{x - n}{r_\beta(n)} )$. By assumption
on $\phi$, $\phi_n \in C^\infty(\RR^d)$ and by assumption on 
$\nn_{\beta, \gamma}$, the $\phi_n$'s 
have disjoint supports. The compactness of the support of $\mu$
gives uniform boundedness of $\omega_n$ in $n$, 
so $V^\omega f \in \WW(\RR^d)$ whenever $f \in \WW(\RR^d)$.  Therefore
if we show that $[\phi_n, A]$, computed on $\WW(\RR^d)$, is bounded for
each $n \in \NN_{\beta, \gamma}$, then in view of the equality 
$$
i[V^\omega, A] = - \sum_{n \in \NN_{\beta, \gamma}} \omega_n \frac{1}{|n|^{\alpha}r_\beta(n)} x\cdot (\nabla \phi)_n(x).
$$ 
it follows that $[V^\omega, A]$ extends to a bounded operator from $\WW(\RR^d)$.
Since $\phi$ is a $C^\infty(\RR^d)$ function of compact support, 
$\nabla \phi$ also has components having compact support and as a consequence 
$\mathrm{supp}(\nabla \phi_n)_j \subset \mathrm{supp} ~ (\phi_n), ~ j=1,\dots,d$
 for all $n \in \nn_{\beta, \gamma}$.
In addition the supports of $\{\phi_n\}$ are mutually disjoint by the
assumption of $\nn_{\beta, \gamma}$, therefore for 
$x \in \mathrm{supp} (\phi_n)$, we have
\begin{equation}\label{eqn03}
\begin{split}
|i[\phi_n, A]f|(x) &\leq |\frac{x}{|n|^{\alpha}r_\beta(n)}\cdot (\nabla \phi)
(\frac{x-n}{r_\beta(n)})||f|(x) \\ 
& \leq 
|\frac{c}{|n|^\alpha}\frac{x-n}{|n|^\beta}\cdot (\nabla \phi)(\frac{x-n}{|n|^\beta})||f|(x)~ 
\\ & ~ ~ ~ + ~ 
| \frac{n}{|n|^{\alpha +\beta}}\cdot (\nabla \phi)(\frac{x-n}{|n|^\beta})|_\infty |f|(x)
\\ & \leq c|(\nabla \phi)|_\infty |f|(x)
\end{split}
\end{equation}
for each $f \in \WW(\RR^d)$. This inequality gives the bound
\begin{equation}\begin{split}
\|i[V^\omega, A]f\|^2 & = \sum_{n \in \nn_{\beta,\gamma}} \int_{\mathrm{supp}(\phi_n)}|\omega_n|^2 |i[\phi_n, A]f|^2(x) ~ dx \\ 
& \leq  c \int_{\mathrm{supp}(\phi_n)} \sup_{n} |\omega_n|^2\|i[\phi_n, A]f\| \leq c \|\nabla \phi\|_\infty \|f\|^2,
\end{split}
\end{equation}
which gives the stated boundedness.

Therefore the commutator $[(H^\omega\pm i)^{-1}, A]$ satisfying the relation 
$$
[(H^\omega\pm i)^{-1}, A] = (H^\omega\pm i)^{-1}[A, H^\omega](H^\omega\pm i)^{-1}
$$
also extends to a bounded operator on $\hh$. Hence    
$$
\|A(H^\omega \pm i)^{-1} f \| \leq 
\|[A, (H^\omega\pm i)^{-1}]f\| + \|(H^\omega \pm i)^{-1} Af\|
$$
implies that $(H^\omega \pm i)^{-1}$ maps $\WW(\RR^d)$ into $D(A)$.  Thus
$\WW(\RR^d)$ is contained in the set 
$$
\{f \in D(A) : (H^\omega \pm i)^{-1} f \in D(A) \}.
$$ 
Since $\WW(\RR^d)$ is a core for $A$, so is the above set, thus we have
verified the conditions (1), (2) of the virial theorem (theorem \ref{thm4}).

Therefore for any normalized eigenvector $f^\omega $ of $H^\omega$, 
we should have
\begin{equation}\label{eqn3}
\langle f^\omega, i[H^\omega, A] f^\omega\rangle = 0.
\end{equation}
However since
\begin{equation}\label{eqn9}
i[H^\omega, A] = 2H^\omega + ([V^\omega, A] - 2 V^\omega) = 2 H^\omega + B^\omega
\end{equation}
with $B^\omega$ bounded  and $ \sup_{\omega} \|B^\omega\| = 2 E_0$ finite,
we see that if $f^\omega$ is the eigen vector of an eigen value $\lambda^\omega$
of $H^\omega$ satisfying $\lambda^\omega > E_0 $, then we must have
\begin{equation}\label{eqn8}
|\langle f^\omega, i[H^\omega, A] f^\omega\rangle|
\geq |\langle f^\omega, (2H^\omega + 2 B^\omega) f^\omega\rangle|
\geq 2 \lambda^\omega  - 2 E_0 > 0,
\end{equation}
contradicting the virial relation given in equation \ref{eqn3} .  
Hence there can be no eigenvalue for $H^\omega_{\beta,\gamma,\alpha}$ bigger than $E_0$. 

To show (ii) we
verify the Mourre estimate in this case.  Let $\chi_I$ denote the indicator function of the
set $I$.  Applying $\chi_{(E_1,\infty)}(H^\omega)$
on either side of equation (\ref{eqn9}) we see that, with $c >0$,
\begin{eqnarray*}
& & \chi_{(E_1,\infty)}(H^\omega)i[H^\omega, A] \chi_{(E_1,\infty)}(H^\omega) \\
& &  > 2(E_1 - \sup_{\omega}\|B^\omega\|) \chi_{(E_1,\infty)}(H^\omega)
> c \chi_{(E_1,\infty)}(H^\omega),
\end{eqnarray*}
for any $E_1 > E_0$ from the inequality (\ref{eqn3}), hence for any closed interval $I$
in $(E_0,\infty)$ we have
$$
\chi_{I}(H^\omega)i[H^\omega, A] \chi_{I}(H^\omega) > c \chi_{I}(H^\omega).
$$
Therefore we only need to verify that the second commutator of $H^\omega$ with respect to $A$ is
relatively bounded with respect to $H^\omega$. Since $\phi$ is smooth
we get that 
$$
i[i[H^\omega, A], A] = 4(H^\omega - V^\omega) - [[V^\omega, A], A]
$$
with
\begin{equation}
\begin{split}
[[V^\omega, A], A] &= \sum_{n\in \NN_\beta} \omega_n (x\cdot \nabla)(x\cdot \nabla)\frac{1}{|n|^\alpha}(\phi)(\frac{x-n}{r_\beta(n)}) \\
 & \approx \sum_{n\in \NN_\beta} \omega_n \left(  \frac{1}{|n|^{\alpha+\beta}}x\cdot 
(\nabla \phi) +  \sum_{j,k=1}^d \frac{x_jx_k}{|n|^{\alpha+2\beta}} 
(\frac{\partial}{\partial x_j} \frac{\partial}{\partial x_k} \phi)(\frac{x-n}{r_\beta(n)}) \right).
\end{split}
\end{equation}
The conditions on $\phi, \alpha, \beta$ are such that the right hand side
is a bounded function of $x$ showing that $[[V^\omega, A],A]$ extends to a
bounded operator. 
Thus 
$i[i[H^\omega, A], A](H^\omega +i)^{-1}$ is bounded.  These estimates
show that the conditions (1)-(5) (taking $K =0, S=I$ there) in definition 3.5.5 
in \cite{DemKri} are satisfied showing that $A$ is a local conjugate 
of $H^\omega$
for each $\omega$.  Hence by Mourre's theorem (theorem 3.5.6 (ii), 
\cite{DemKri})
there is no singular continuous spectrum for $H^\omega$
in $I$.  These two results together show that there is no singular spectrum
in any closed subinterval of $(E_0, \infty)$, showing the theorem.  \qed

{\noindent \bf Proof of theorem \ref{cor1}:}

When $\alpha >0$, the potential is relatively compact with respect to $-\Delta$,
so Weyl's theorem implies the statement on the essential spectrum.  
On the other hand since
$H^\omega_{\beta,\gamma,\alpha}$ is an unbounded self adjoint operator
its spectrum cannot be bounded hence the statement for $\alpha =0$. \qed

{\noindent \bf Proof of Theorem \ref{thm5}:}

(ii) The proof is almost as in the proof of Theorem 3.5 of 
Hislop-Kirsch-Krishna \cite{hkk} with a minor modification.
The equation (17) of \cite{hkk} should be replaced by
\begin{equation}
\begin{split}
P_\n ( H_\Lambda^\omega - E - i \epsilon )^{-1} P_\m & =  P_\n
 ( H_0 - E - i \epsilon )^{-1} P_\m
 \nonumber \\
 & ~  - \sum_{\k \in I_\Lambda} P_\n ( H_0 - E - i \epsilon )^{-1} \omega_\k
 P_\k  ( H_\Lambda^\omega - E - i \epsilon )^{-1} P_\m .
 \nonumber \\
\end{split}
\end{equation}
Then the estimate in the inequality (21) of \cite{hkk} should be redone as
\begin{equation}
\begin{split}
\lefteqn{ \EE \{ \| P_\n ( H_\Lambda^\omega - z)^{-1} P_\m \|^s \} } \nonumber \\
 & \leq  \| P_\n ( H_0 - z )^{-1} P_\m \|^s  \nonumber \\
 &  + K_s \; \sum_{\k \in I_\Lambda} \| P_\n ( H_0 - z )^{-1} P_\k \|^s \; \EE \{ \| P_k
      ( H_\Lambda^\omega - z )^{-1} P_\m \|^s \} \\
 & \leq  \| P_\n ( H_0 - z )^{-1} P_\m \|^s  \nonumber \\
 &  + K_s \; \sum_{\k \in I} \| P_\n ( H_0 - z )^{-1} P_\k \|^s \; \EE \{ \| P_k
      ( H_\Lambda^\omega - z )^{-1} P_\m \|^s \} .
\end{split}
\end{equation}
Now the proof goes through exactly as that of Theorem 3.5 Hislop-Kirsch-Krishna \cite{hkk}.

(i) To show that the a.c. spectrum contains $[0, \infty)$
we  prove that wave operators
for the pair $(H^\omega_1, -\Delta)$ exist almost everywhere. 

The existence of wave operators follows if we show that for a dense
set of $f \in L^2(\RR^d)$, the limits
$$
\lim e^{iH^\omega_1 t} e^{i\Delta t} f
$$
exist strongly as $t$ goes to $\infty$.  Thus by cook's method  
$$
\lim_{s,t \rightarrow \infty} \|e^{iH_1^\omega t}e^{i\Delta t} - e^{iH_1^\omega s} e^{i\Delta s}f \| 
\leq \lim_{s,t \rightarrow \infty}
 \int_s^t dw \|(H_1^\omega + \Delta) e^{i\Delta w}f \|  = 0  
$$
for a dense set of $f$.  This follows if the integral 
\begin{equation}\label{eqn010}
\int_1^\infty dt ~ \|V^\omega e^{i\Delta t}f \| < \infty,
\end{equation}
for a dense set of $f$.

Let the union of the coordinate axes in $\RR^d$ be denoted by $A_0$,
thus $A_0 = \{x\in \RR^d : x_i = 0 ~ \mathrm{for ~ some } ~ i = 1,\dots,d \}$.
We pick the dense set to be 
$$
\dd = \{f \in L^2(\RR^d) : supp ~ \widehat{f}  
\subset \RR^d \setminus A_0 ~ \mathrm{and ~ supp ~ \widehat{f} ~ compact} \}. 
$$
We therefore consider the integrand and get the estimate for each $\omega$,
\begin{equation}\label{eqn10}
\begin{split}
\|V^\omega e^{i\Delta t} f\| & = \|\sum_{\n \in I_\Lambda }\omega_n 
\langle \Phi_\n, e^{i\Delta t} f\rangle \Phi_\n\| \\
& \leq C \sum_{\n \in I_\Lambda } |\langle \Phi_\n, e^{i\Delta t} f\rangle |
\end{split}
\end{equation}
since $\omega_n$ are bounded and $\{\Phi_\n \}$ is an orthonormal set.
We will show that the sum in inequality ( \ref{eqn10}) converges.

We first note that under taking Fourier transforms we have 
\begin{equation}
\label{eqn11}
\begin{split}
\langle \Phi_\n, e^{i\Delta t} f\rangle & = \int \widehat{\overline{\Phi_\n}}
(\xi)e^{-i\xi^2 t } \widehat{f}(\xi) ~ d\xi \\
& = 2^{-\frac{dn_1}{2}}\int \widehat{\overline{\Psi_{c(\n)}(\xi)}}~  
e^{-i2^{-n_1} n_2\cdot \xi-i\xi^2 t } \widehat{f}(\xi) ~ d\xi. 
\end{split}
\end{equation}

We recall (from equations (\ref{eqn100}, \ref{eqn101}) ) that the
function $\widehat{\Psi_{c(\n)}}$ has at least one factor $\psi_j$ 
(which is supported in the set $\{|\xi_j| \in [2\pi/3, 8\pi/3] \}$)
so that for at least one coordinate $\xi_j$ of $\xi$ we have the
condition $|2^{-n_1}\xi_j| \in [2\pi/3, 8\pi/3]$.  In addition by
the choice of $f$ we have $\widehat{f}(\xi) = 0 $ if $|\xi_j| \notin [c, d]$
for some $0 < c < d < \infty$ for all $j=1,\dots, d$.  These two conditions 
together imply that the integral is zero unless there is an $R < \infty$
such that $ - R < n_1 < R$, where $R$ depends on $c,d$.  Thus the sum
over $n_1$ is reduced to a finite sum in equation (\ref{eqn10}).  

The idea is now to get arbitrary decay in $t$ from the integral with respect to $\xi_1$
and get decay in each of the variables ${n_2}_j$ in exchange for some growth in $t$ from each
of the other variables $\xi_2, \dots , \xi_d$.   These estimates together give decay of the integral
in both $t$ and $|n|$.  

By assumption on $\Lambda$, writing $n_2 = ({n_{2}}_1, \dots, {n_{2}}_d)$,  ${n_{2}}_k$ is finite
for some $k = 1, \dots d$, without loss of generality let $|{n_{2}}_1 | < K < \infty$.
We set $a_1(\xi_1) = -i \frac{\partial}{\partial \xi_1} (2^{n_1}{n_{2}}_1\xi_1 - \xi_1^2 t)$. Then we have
\begin{equation}\label{eqn12}
|a_1(\xi_1)| = |\frac{\partial}{\partial \xi_1} (2^{n_1}{n_{2}}_1\xi_1 - \xi_1^2 t) |
\geq 2 |t| \vline |\xi_1| - \frac{2^{n_1-1}{n_{2}}_1 }{t} \vline \geq 2 |t| c/2 = c |t|, 
\end{equation}
if $2^{R-1}K /t < c/2$, whenever $\xi \in \mathrm{supp}{\widehat{f}}$.  Under the 
hypothesis \ref{hyp1.1}, $\widehat{\Psi_\n}$ has $2d+2$ partial derivatives
in each of the $\xi_j$'s, so
we can do repeated integration by parts with respect to the variable $\xi_1$
in the above integral equation (\ref{eqn11}) to get, for every $\ell \in \{1,2,\dots, 2d+2\}$,
\begin{equation}\label{eqn13}
\begin{split}
\langle \Phi_\n, e^{i\Delta t} f\rangle 
& = (-1)^\ell \int e^{-i2^{-n_1} n_2\cdot \xi-i\xi^2 t } ~ \left(\frac{\partial}{\partial \xi_1} \frac{1}{a_1(\xi_1)}\right)^\ell B_\n(\xi) ~ d\xi, 
\end{split}
\end{equation}   
where we took $2^{-\frac{dn_1}{2}}\overline{\widehat{\Psi_{c(\n)}}}(2^{-n_1} \xi) \widehat{f}(\xi) = B(\n, \xi)$.

We now take 
\begin{equation}\label{eqn14}
B(t,\ell, \n, \xi) = e^{-i\xi^2t}\left(\frac{\partial}{\partial \xi_1} \frac{1}{a_1(\xi_1)}\right)^\ell B_\n(\xi)
\end{equation}
and do integration by parts twice with respect to each of the variables 
$\xi_2, \dots, \xi_d$ to get
\begin{equation}\label{eqn15}
\begin{split}
\langle \Phi_\n, e^{i\Delta t} f\rangle 
& = \left(\prod_{j=2}^d \frac{-1}{2^{-2n_1}{n_{2}}_j^2 }\right)(-1)^\ell \int e^{-i2^{-n_1} n_2\cdot \xi} ~ \prod_{j=2}^d \frac{\partial^2}{\partial \xi_j^2} B(t,\ell,\n,\xi) ~ d\xi.
\end{split}
\end{equation} 

It is now a tedious but not difficult calculation to see, using 
inequalities/equations (\ref{eqn12} - \ref{eqn15}),
that if we take $\ell = 2d$, then
$$
|\langle \Phi_\n, e^{i\Delta t} f\rangle| \leq C \prod_{j=2}^d \frac{1}{1+ |{n_{2}}_j|^2} |t|^{2d-2}{|t|^{-2d}} \int W(n_1, \xi) ~ d\xi,
$$
where the factor $|t|^{2d-2}$ is the maximum power of $|t|$
possible by taking derivatives of the factor
$e^{-i\xi^2t}$ with respect to the variables $\xi_2, \dots, \xi_d$, while
the factor $|t|^{-\ell}$ comes from the factor $1/a_1(t, \xi)$ occurring
$\ell$ times. 
and we clubbed all the rest of the integrand in $W$. Using the fact
that $|n_1| < R$ and that $W$ has compact support in $\xi$ and so is
integrable, the above inequality implies that 
$$
\int_1^\infty \sum_{\n \in \Lambda} |\langle \Phi_\n, e^{i\Delta t} f\rangle|
 ~ dt <  C \int_1^\infty t^{-2} ~ dt \sum_{ |n_1| <\infty}\sum_{{n_{2}}_2,\dots, {n_{2}}_d \in \ZZ} \frac{1}{1+ |{n_{2}}_j|^2} < \infty.
$$ 
This estimate together with the inequality (\ref{eqn10}) proves the
required inequality (\ref{eqn010}). \qed

\section{Examples}

There are lots of examples of sets $\NN_{\beta, \gamma}$ mentioned
before equation (\ref{pot1}).

\begin{exam}\label{exam1}
We shall give an example in $d=2$ of the potentials that have neither 
'decay' nor supported on a 'sparse' set.  This example is motivated
by the paper of Rodnianski-Schlag \cite{RodSch}.

Consider a fixed $R >0$ and consider the squares $B_k = \{ x \in \RR^2 :
|x_i|\leq 2^k R, i=1,2 \}, ~ k \in \ZZ^+$, which are centered at the
origin and have side length $2^{k+1}R$.  Then $\cup_{k} B_k = \RR^2$
and we consider the annulus $A_k = B_{k+1} \setminus B_k$. The area 
of $B_k$ is $(2^{k+1}R)^2$ and so the area of the annulus is
$Area(A_k) = Area(B_{k+1}) - Area(B_k) = 3 (2^{k+1}R)^2$.  Clearly
we can cover $A_k$ with 12 squares of side length $2^kR $ each, with
the centres of these squares falling on the lines $|x_1|= 3 2^{k-1} R$
or $|x_2| = 3 2^{k-1}R$.  We take the squares $S_y$ of side length $2^{k}R$
centred at the points $y$ in the set  
\begin{equation}
\begin{split}
C_k & = \{ x : x_2 = \pm \frac{3}{2} \times 2^k R, x_1= \pm 2^{k-1}R, \pm 3 \times  2^{k-1}R \}
\\ & ~ ~ \bigcup \{x : x_1 = \pm \frac{3}{2} \times 2^k R, x_2= \pm 2^{k-1}R, \pm 3 \times  2^{k-1}R \}
\end{split}
\end{equation}   
and take the respective discs of radius $2^{k-1}R$ with the same
centres and inscribed in the squares.  We can then take a bump function
$\phi$ supported in the unit disk, nowhere vanishing in the open disk but
vanishing on its boundary. We take $r_1(n) = 2^{k-1}R$ for $n \in C_k$.
Since the points of $C_k$ have absolute value $\sqrt{10} 2^{k-1}R$ or $\sqrt{18} 2^{k-1}R$, we find that the condition
$$
\frac{1}{3\sqrt{2}} |n| \leq r_1(n) = 2^{k-1}R \leq \frac{1}{\sqrt{10}}|n|, ~ n \in C_k
$$
is valid.
Then the functions 
$\phi(\frac{x-y}{2^{k-1}R})$
with $y \in C_k$ give a collection of functions such that 
$$
\sup_{k \in \ZZ^+} \sup_{x \in \RR^2} |(x.\nabla \phi)(\frac{x-y}{2^{k-1}R})| < \infty ~~ \mathrm{and} ~~ 
\sup_{k \in \ZZ^+} \sup_{x \in \RR^2} |(x.\nabla \phi)^2(\frac{x-y}{2^{k-1}R})| < \infty.
$$    
Further we note that by construction, for each $k$ we have
$$
Area(A_k) = \cup_{y \in C_k} Area(S_y) = 12 (2^kR)^2.
$$
The area of the discs inscribed
in each $S_y , y \in C_k$ is $ \pi (2^{k-1}R)^2$, so the total area
of these discs contained in $A_k$ is $12 \pi (2^{k-1}R)^2$.  Thus
the in each of the annuli $A_k$ the area of the discs is $\frac{\pi}{4} Area(A_k)$.  Adding up we find that the union of the discs we constructed
with centers at all points in $C_k$ is a fraction $\frac{\pi}{4}$, so 
it also forms the same fraction of the area of the squares $B_k$ .
This shows that the union of the supports of the
functions $\phi(\frac{x-y}{2^{k-1}R}), y \in C_k$ has positive density in 
$\RR^2$ (the density of $\frac{\pi}{4}$).  
\end{exam}
\begin{rem}
\begin{enumerate}
\item It is clear from the construction above example that 
if we took a product of bump functions $\prod_{j=1}^d f_j$
each supported on $[-1,1]$, then we can get $\phi_y$'s
to have full support in the annulus $A_k$ and then the
resulting potential  
$$
V^\omega(x) = \sum_{n \in \cup_{k=1}^\infty C_k } \omega_n \phi_n(x)
$$
is a random potential which is non-vanishing on a set of full measure
on $\RR^d$ for which  Theorem \ref{thm3} will be valid. Of course
there are many more possibilities.

\item The above example can be extended to any $\RR^d$ with spheres replacing
discs, but the centers chosen to fall in between cubes $\Lambda_k$ 
of side lengths $2 \gamma^k R$, $\gamma > 1$ centred at the origin. 
The spheres can be chosen to lie in the region $\Lambda_{k+1} \setminus
\Lambda_k$ with centres chosen so they pack a positive density (which
is independent of $k$ but depends on the dimension $d$) of the volume
of this region.  Such sets give rise to independent random potentials 
(which are supported on these sets ) that
are neither 'decaying' nor have 'sparse' supports.  Nevertheless
there is no localization at large energies for them. 
\end{enumerate}
\end{rem}

\section{Appendix}

We reproduce verbatim the construction of the projections $P_\n$,
used in equation \ref{theothermodel}, 
using the Lemare-Meyer wavelets from 
Hislop-Kirsch-Krishna \cite{hkk}, for easy reference.

In order to construct the projections
$P_\n$, we first recall the definition of wavelets in higher dimensions.

A {\it wavelet} in one dimension is a function $\psi$ with the
property that the collection of translated and diadically dilated functions
$\{ \psi_{j,k}(x) = 2^{j/2} \psi(2^{j}x - k) \; | \;  j, k \in \ZZ\}$, 
forms an orthonormal basis for $L^2(\RR)$.
Associated with the wavelet $\psi$ is the {\it scaling function} $\phi$.
The scaling function $\phi$ is used to construct the wavelet $\psi$
through a procedure called {\it multiresolution analysis} (cf.
\cite{[D],[Wo]}).

To define a wavelet in higher dimensions (as in \cite{[Wo]}, Proposition
5.2), we first start with
a collection $\{\phi_1, \dots, \phi_d, \psi_1, \dots, \psi_d\}$
of $2d$ functions on $\RR$ of which the $\phi_j$ are scaling functions and
the $\psi_j$ are the associated wavelets constructed from the $\phi_j$.
We note that we may take all the $\phi_j \equiv \phi, ~~ \psi_j \equiv
\psi, ~~ j=1, \dots, d.$, although this is not necessary. 
Let us define an index set
$F = \{ c = (c_1 , \ldots , c_d ) \in \{0,1\}^d \setminus (0,0,\dots, 0) \}$.
For each $c \in F$, we define a function on $\RR^d$ by   
\begin{equation}
\label{eqn100}
\Psi_c (x) = \prod_{j=1}^d (\delta_{c_j, 0} \phi_j + \delta_{c_j, 1}
\psi_j)
(x_j), ~~ c \in F.
\end{equation}
Here, the $\delta_{c_j , k}$, for $c = ( c_1 , \ldots , c_d ) \in F$
and $k = 0 , 1$, is the Kronecker delta. In the product, the
function $\phi_j (x_j)$ is present if the index $c_j$ is zero, 
and $\psi_j (x_j)$ is present otherwise.
Note that there is at least one factor $\psi_j$ in $\Psi_c$ for any $c \in
F$. We consider the set of dyadic dilations and 
$\ZZ^d$-lattice translations of these functions.
We denote by $I$ the countable index
set $I = F \times \ZZ \times \ZZ^d$. An element
$\n \in I$ is a triple $\n = ( c(\n) , n_1 , n_2 )$.
The collection of dilated and translated functions
\begin{equation}\label{eqn101}
\Phi_\n ( x) = 2^{n_1 d/2} \Psi_{c(\n)}(2^{n_1}  x - n_2 ), ~~ c \in F,
~~ n_1 \in \ZZ ~~ n_2 \in \ZZ^d, ~~ x \in \RR^d,
\end{equation}  
is called a {\it multi-variable wavelet}
if the collection forms an orthonormal basis for $L^2(\RR^d)$.

In the following we
shall, notationally, always refer to the collection of functions $\{\Psi_c
\; | \; c \in F\}$ simply as $\Psi$, and any property stated for $\Psi$
is by definition to be take to be valid for each member of this collection.
Thus a statement that the property P is valid for $\widehat{\Psi}$ means
that P is valid for each of the Fourier
transforms $\widehat{\Psi_c}$, for each $c \in F$, and so on.

We assume the following conditions on the 
multi-variable wavelet  and the distribution
of the random variables $\{ \omega_n \; | \; \n \in I \}$.
\begin{hyp}
\label{hyp1.1}
Let $\Psi$ be a multi-variable wavelet formed out of the scaling functions
$\phi_i, i = 1, \dots, d$ and the wavelets $\psi_i, i=1, \dots, d$ such
that
\begin{enumerate}
\item the functions $\widehat{\phi_j} \in \cc^{2d+2}(\RR), \widehat{\psi_j} \in \cc^{2d+2}_0(\RR), ~~ j = 1, \dots, d$;
\item the functions $\widehat{\phi}_j^{(\alpha)}$, 
for $|\alpha | \leq 2d+2 $, decay rapidly;
\item the functions are normalized, $\int | \Psi |^2 ~dx = 1$.
\end{enumerate}
\end{hyp}

\begin{hyp}
\label{hyp1.2}
Let $I = F \times \ZZ \times \ZZ^d$, and
let $\{ \omega_n \; | \;  \n \in I\}$ be independent and 
identically distributed random variables with their common 
probability distribution $\mu$ being
absolutely continuous and of compact support in $\RR$.
\end{hyp}

{\noindent \bf Remarks:} 
Any one-dimensional Lemari\'e-Meyer wavelet
$\psi$, and its related scaling function $\phi$, satisfy Hypothesis
\ref{hyp1.1}. Typically, a Meyer wavelet can be constructed to be in the
Schwartz class, $\psi \in {\cal S} ( \RR)$, and its
Fourier transform $\hat{ \psi}$ is compactly
supported in the set $[ - 8 \pi / 3 , - 2 \pi / 3] \cup 
[ 2 \pi / 3 , 8 \pi / 3 ]$. The corresponding scaling
function can also be chosen to satisfy
$\phi \in {\cal S} ( \RR)$, and so that $\hat{ \phi}$ has compact
support in $[- 4 \pi / 3 , 4 \pi / 3]$, cf.\ \cite{[LM],[Wo]}.
A large number of additional examples are 
constructed in the paper of Auscher, Weiss, and Wickerhauser \cite{[AWW]}.

\end{document}